\documentstyle[twoside]{article}

\input pz.sty
\input epsf.sty

\begin{document}

\PZtitletl{Spectral Evidence for Appearance of New Decretion Disk Around IGR J06074+2205}{ }

\PZauth{A.~O.~Simon$^{1}$, A.~V.~Bondar$^{2}$, N.~V.~Metlova$^{3}$}

\PZinst{Taras Shevchenko National University of Kyiv, Glushkova ave. 4,
03127 Kyiv, Ukraine; e-mail: andrew\_simon@mail.ru}

\PZinst{International Center for Astronomical, Medical and Ecological Research Terskol Observatory}

\PZinst{Crimean Laboratory of Sternberg Astronomical Institute, Nauchny, Crimea, Ukraine}

\PZabstract{We report about new episode of decretion disk formation in the IGR J06074+2205 system. Obtained spectral data gives us opportunity to measure peak separation in double-peaked H$_{\alpha}$ line as 408$\pm$55 km/s and hence obtain disk radii as 1.6 star radii. All these facts may says about possible X-ray activity of IGR J06074+2205 in the nearest future.}

\PZbegintext{A.O.Simon, A.V.Bondar, N.V.Metlova: New disk around IGR J06074+2205}

\section{{\large \bf Introduction}}

IGR J06074+2205 is a NS Be/X-ray binary of B0.5V spectral type. The distance to the object is about 4.5 kpc. From investigation of three HeI lines the rotational velocity of the star was determined as 260$\pm$20 km/s. And after the five years of the spectral monitoring of IGR J06074+2205 (P.Reig et al., 2010) it was reported about the fact of reversion from emission to absorption of the H$_{\alpha}$ line and hence disappearance of the equatorial disk. The total disk disappearance was observed in March 2010 when H$_{\alpha}$ line profile fully have gone into absorption.

In general, disk appearance and disappearance episodes are usual for Be/X-ray binaries but disk loss episodes are more observed. Now it is known that decretion disks in these systems exists about of 3-7 years (see Table 5, P.Reig, 2011) and typical disk loss episodes lasted about 1.5-2 years. Due to absence of significant amount of information about disk producing and growing up episodes we put the aim to study disks in the phase of its formation and IGR J06074+2205 can be the first candidate for this study.

\section{{\large \bf Observations and Data Reduction}}

IGR J06074+2205 was observed April 14 and April 24, 2012 at International Center for Astronomical, Medical and Ecological Research with the 2m Ritchey-Chretein-Coude telescope with Cassegrain Multi Mode Spectrograph (CMMS) (with R = 14000) as a part of spectral monitoring of the selected Be/X-ray binaries. We have obtained one spectra per night with 2700$^s$ exposure (A.Simon et al., 2012) for the object. Calibration frames were taken the same nights. For our analysis we have used only one order with H$_{\alpha}$ line from the whole echelle spectrum. The SNR value in the H$_{\alpha}$ region is 10.

All data were proceed with Dech95 software (G.A. Galazutdinov). Due to such low SNR value it was impossible to use bias and dark frames for calibration due to presence of negative pixel's intensities in the resulting frame. So we decide to subtract from the selected echelle order the noise measured both sides of this order. The subtracted spectral orders were analyzed with Dech20Tsoftware (G.A. Galazutdinov). To make a dispersion curve we have used both Fe-Ar lamp specter and day sky specter. With the help of dispersion curve we have determine the spectral dispersion as 0.25 {\AA}/px.

\section{{\large \bf Data Analysis and Results}}

Using our spectrum of IGR J06074+2205 obtained April 14 (black line on Fig.1) and April 24 (red line on Fig.1) we can clearly identify double peek H$_{\alpha}$ emission line and its V/R variability. Appearance of H$_{\alpha}$ emission line may indicate about appearance of the equatorial disk around Be star. In spite of high level noise spectra it is possible to conclude that symmetric H$_{\alpha}$ line has a shell profile with FWHM=12.4$\pm$0.5{\AA}. Peaks separations for this line is 408$\pm$55 km/s. Taking into account the star's rotational velocity 260$\pm$20 km/s (P.Reig et al., 2010) and correlation between star's and disc radii one can obtain the disk radii of about 1.6 star radii.

\PZfig{8cm}{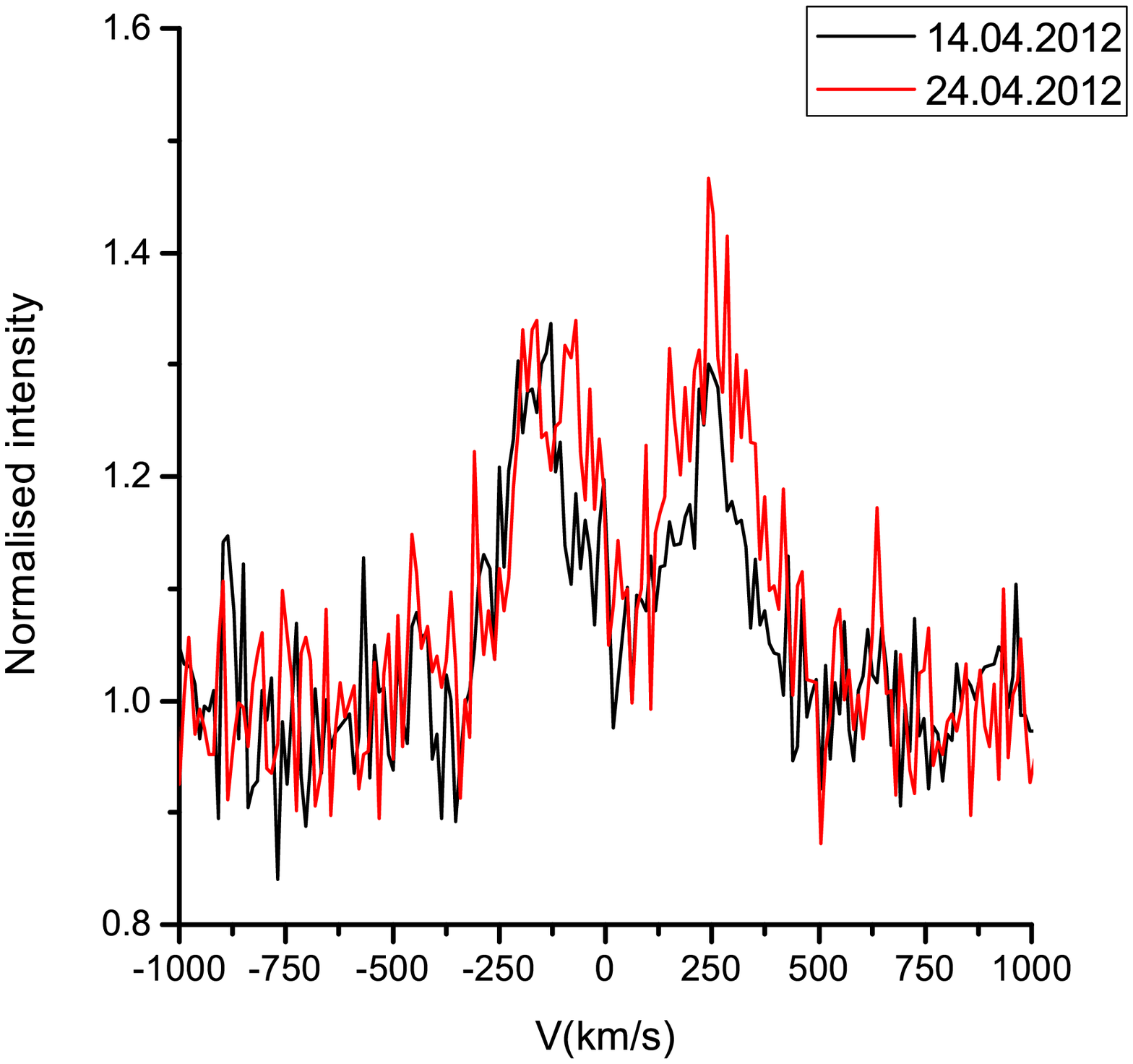}{Two samples of H$_{\alpha}$ line profiles obtained 2012 April, 14  - black and April, 24 - red}

\section{{\large \bf Discussion}}

The timescale between full disk disappearance (H$_{\alpha}$ in absorption) and its next appearance (H$_{\alpha}$ in emission again) for the case of IGR J06074+2205 is no more than 26 months (or about 2 years). This timescale may be smaller due to absence of observations during this two years. For example, in some other objects the time between observed last absorption and firs emission in H$_{\alpha}$ are believed to be about some months. In the case of A0535+26 (see Fig.2, N.J. Haigh et al., 1999) such timescale is about 3 month. For V635 Cas (I. Negueruela et al., 2001) it was obtained only 2 spectra with the time interval of 4 months which shows changes from absorbtion to emission. But there are some other systems, such as X Per (J.S. Clark et al., 2001) and RX J0440.9+4431 (P.Reig et al., 2005), which didn't show any absorbtion lines during disk loss episodes. Their H$_{\alpha}$ profiles were mostly flat like continuum (X Per) or very weak emission (EW$\le$1).

Disk appearance and its small size (1.6 star radii) gives us opportunity to predict a new episode of optical and possible X-ray activity of IGR J06074+2205 in the nearest future with its V/R variability and increasing of IR excess. Further spectral and photometrical observations in the optical and IR bands are encouraged and useful for understanding disk formation mechanisms.

\bigskip

{\bf Acknowledgments:} This work was supported by the International program "Astronomy in the Elbrus region (2010-2014)"
of ICAMER. We thanks the stuff of the ICAMER for the help in observations.

\references

J.S. Clark, A.E. Tarasov, A.T. Okazaki et al., 2001, {\it A\&A}, {\bf 380}, 615

G.A. Galazutdinov, http://boao.re.kr/gala/dech.htm

N.J. Haigh; M.J. Coe; I.A. Steele; J. Fabregat, 1999, {\it MNRAS}, {\bf 310}, L21

I. Negueruela; A.T. Okazaki, 2001, {\it A\&A}, {\bf 369}, 108

P.Reig, I. Negueruela, J. Fabergat et al., 2005, {\it A\&A}, {\bf 440}, 1079

P. Reig, A. Zezas, L. Gkouvelis, 2010, {\it A\&A}, {\bf 522}, A107

P. Reig, 2011, {\it Ap\&SS}, {\bf 332}, 1

A. Simon, 2012, {\it ATel}, {\bf 4172}

\endreferences

\end{document}